\documentclass[lettersize,journal]{IEEEtran}
\usepackage{amsmath,amsfonts}
\usepackage{algorithm}
\usepackage{array}
\usepackage[caption=false,font=normalsize,labelfont=sf,textfont=sf]{subfig}
\usepackage{textcomp}
\usepackage{stfloats}
\usepackage{url}
\usepackage{verbatim}
\usepackage{graphicx}
\usepackage{booktabs}
\usepackage{multirow}
\usepackage{adjustbox} 
\usepackage{algpseudocode}
\usepackage[table]{xcolor}
\usepackage{mathtools}
\usepackage{algorithm}      
\usepackage[table]{xcolor}
\algtext*{EndIf} 
\algtext*{EndFor} 
\usepackage[backend=biber,style=ieee,sorting=none,
  doi=false,isbn=false,url=false,eprint=false]{biblatex}
\begin{document}

\newcommand{\jianhua}[1]{\textcolor{red}{XXX Jianhua: #1}} 
\newcommand{\zhouyixiang}[1]{\textcolor{blue}{zhouyixiang: #1}} 
\title{Fed-PELAD: Communication-Efficient Federated Learning for Massive MIMO CSI Feedback with Personalized Encoders and a LoRA-Adapted Shared Decoder\\}

\author{
 \IEEEauthorblockN{
        Yixiang Zhou,
        Tong Wu,
        Meixia Tao, and
        Jianhua Mo
    } \\
 \thanks{The authors are with the School of Information Science and Electronic Engineering, Shanghai Jiao Tong University, China (e-mail: yxzhou02@163.com, \{w69698, mxtao, mjh\}@sjtu.edu.cn). Corresponding authors: Meixia Tao and Jianhua Mo.}
}



\maketitle

\begin{abstract}
This paper addresses the critical challenges of communication overhead, data heterogeneity, and privacy in deep learning for channel state information (CSI) feedback in massive MIMO systems. To this end, we propose Fed-PELAD, a novel federated learning framework that incorporates personalized encoders and a LoRA-adapted shared decoder. Specifically, personalized encoders are trained locally on each user equipment (UE) to capture device-specific channel characteristics, while a shared decoder is updated globally via the coordination of the base station (BS) by using Low-Rank Adaptation (LoRA). This design ensures that only compact LoRA adapter parameters instead of full model updates are transmitted for aggregation. To further enhance convergence stability, we introduce an alternating freezing strategy with calibrated learning-rate ratio during LoRA aggregation. Extensive simulations on 3GPP-standard channel models demonstrate that Fed-PELAD requires only 42.97\%  of the uplink communication cost compared to conventional methods while achieving a performance gain of 1.2 dB in CSI feedback accuracy under heterogeneous conditions.

\end{abstract}


\section{Introduction}
\IEEEPARstart{M}{assive} multiple-input multiple-output (MIMO) technology is a cornerstone of 5G and a key enabler for 6G, where base stations (BS) deploy extensive antenna arrays to enhance area spectral efficiency significantly \cite{Alkhateeb2014MIMO}. Harnessing these benefits depends critically on the availability of timely and accurate channel state information (CSI) at both sthe transmitter and receiver. In frequency-division duplexing (FDD) systems, the downlink CSI must be estimated at user equipments (UEs) and fed back to the BS via the uplink channel, incurring a signaling overhead that escalates dramatically with the increasing number of BS antennas and the demanded feedback resolution.

Conventional methods on CSI feedback in FDD massive MIMO systems mainly include codebook-based schemes \cite{ziao2025review} and compresssive sensing-based schemes \cite{gao2018compressive}. The former quantizes CSI into predefined codewords but typically suffers from high search complexity and substantial feedback overhead. The latter compresses CSI via random projections, but involves computationally expensive iterative reconstruction. The advent of deep learning (DL) has ushered in a transformative paradigm for CSI feedback. Models employing convolutional neural networks (CNNs) \cite{wen2018deep} or transformer architectures \cite{cui2022transnet} have demonstrated remarkable improvements in reconstruction accuracy under sophisticated channel conditions, positioning them as a key focus within 3GPP standardization efforts \cite{3gpp38843v19}. 

Most existing DL-based CSI feedback frameworks rely on centralized model training, which aggregates raw CSI data from distributed UEs. This approach raises two key challenges. First, collecting large-scale CSI data from distributed UEs may compromise user privacy, as CSI inherently encodes sensitive mobility and location information. Second, CSI distributions across UEs are inherently heterogeneous due to environmental and hardware diversity, leading to severe performance degradation when a single global model is trained centrally. These challenges motivate the exploration of distributed learning of CSI feedback among UEs that can preserve data locality and handle data heterogeneity.

Federated edge learning (FEEL) has emerged as a promising distributed learning framework that can address the data privacy issue by enabling collaborative model training across UEs without sharing raw data \cite{tao2024federated}. In FEEL, each UE trains the global model using its own local data and sends the model updates to the BS for aggregation iteratively. The work \cite{cui2024personalizedfededge} has applied FEEL to train end-to-end CSI feedback autoencoders as global models, subsequently refining them through UE-specific fine-tuning to accommodate heterogeneous local data distributions. While this personalization strategy partially addresses distribution shifts across UEs, it introduces substantial communication overhead inherent in FEEL for exchanging model parameters. Moreover, since the CSI feedback decoders are also personalized for each UE, it brings increasing storage burden at the BS when the number of UEs increases. 
To address these issues, FedDec \cite{du2023training} employs private encoders and aggregates only decoder parameters, while CSILocal \cite{dong2025modelsplitting} leverages model splitting and pipeline parallelism. These efforts can achieve a degree of communication reduction, but still lack dedicated mechanisms to ensure model robustness and generalization under the prevalent heterogeneous data distributions across UEs.

To overcome the above limitations, we propose Fed-PELAD, a novel \textbf{Fed}erated edge learning framework that incorporates \textbf{P}ersonalized \textbf{E}ncoders and a \textbf{L}oRA-\textbf{A}dapted shared \textbf{D}ecoder for massive MIMO CSI feedback. This learning framework can simultaneously address the challenges of communication efficiency, data privacy, and data heterogeneity with stable convergence. Specifically, the personalized encoders are trained locally on each UE to capture device-specific channel characteristics and eliminate uplink transmission of encoder parameters. 
The LoRA-adapted shared decoder is trained globally across UEs by integrating Low-Rank Adaptation (LoRA), a parameter-efficient fine-tuning technique pioneered for large language model \cite{hu2022lora}. This approach enables substantial reduction in parameter transmission for federated aggregation at the BS by communicating only compact adapter weights instead of full model parameters. 
Furthermore, an alternating-freeze strategy with learning-rate ratio calibration during LoRA aggregation is designed to stabilize convergence under heterogeneous CSI distributions.

Extensive simulations based on 3GPP-standard channel models demonstrate that the proposed Fed-PELAD reduces the uplink communication cost to only 42.97\%  of that required by conventional federated learning, while achieving a performance gain of 1.2 dB in CSI feedback accuracy under heterogeneous channel environment.

\section{System Model}
We consider a single-cell FDD massive MIMO downlink with a BS equipped with
$N_t$ antennas and multiple single-antenna UEs. 
Let $N_s$ denote the number of orthogonal frequency division multiplexing (OFDM) subcarriers. The $n$-th subcarrier channel vector is
$\mathbf h_n\!\in\!\mathbb C^{N_t\times 1}$, and stacking all subcarriers yields
the frequency-domain CSI matrix
\begin{equation}
\tilde{\mathbf H}
=\big[\mathbf h_1,\,\mathbf h_2,\,\ldots,\,\mathbf h_{N_s}\big]^{\!H}
\in\mathbb C^{N_s\times N_t}.
\label{eq:sys-Htilde}
\end{equation}

Exploiting the sparsity of CSI in the angle–delay domain, we apply a 2-D discrete Fourier transform (DFT) to obtain
\begin{equation}
\bar{\mathbf H}=\mathbf F_d\,\tilde{\mathbf H}\,\mathbf F_a^{H}, \label{eq:sys-2ddft}
\end{equation}
where $\mathbf F_d \in \mathbb C^{N_s\times N_s}$,$\mathbf F_a \in \mathbb C^{N_t\times N_t}$ are DFT matrices. We then retain the $N_a$ dominant delay taps ($N_a\!\le\!N_s$) to obtain the cropped
CSI as
\begin{equation}
 \Tilde{\mathbf H}=\left[ \bar{\mathbf H} \right ]_{1:N_a,:}\in\mathbb C^{N_a\times N_t}.
\label{eq:sys-crop}
\end{equation}

Since $\Tilde{\mathbf H}$ is complex-valued, we adopt a real-valued tensor representation $\mathbf H_a \in\mathbb R^{N_a\times N_t \times 2} $ for learning by stacking the real and imaginary parts of $\Tilde{\mathbf H}$.
For each UE, an encoder $f_{\mathrm{en}}(\cdot;\Theta_{\mathrm{en},k})$ compresses
$\mathbf H_a$ into a length-$M$ codeword
\begin{equation}
\mathbf v=f_{\mathrm{en}}(\mathbf H_a;\Theta_{\mathrm{en},k})\in\mathbb R^{M},
\label{eq:sys-enc}
\end{equation}
where $\Theta_{\mathrm{en},k}$ are the encoder parameters of UE $k$.
The compression ratio (CR) is defined as $\gamma=M/(2 N_a N_t)$. The codeword is transmitted via the uplink  channel, and we assume that this process is lossless. After the BS receives the codeword $\mathbf v$, a decoder $f_{\mathrm{de}}(\cdot;\Theta_{\mathrm{de}})$ reconstructs the
cropped CSI,
\begin{equation}
\widehat{\mathbf H}_a=f_{\mathrm{de}}(\mathbf v; \Theta_{\mathrm{de}})
\in\mathbb C^{N_a\times N_t},
\label{eq:sys-dec}
\end{equation}
where $\Theta_{\mathrm{de}}$ denotes the decoder parameters.

The loss function is the normalized mean squared error (NMSE), 
\begin{equation}
\mathcal L(\Theta_{\mathrm{en},k},\Theta_{\mathrm{de}})
=\mathbb E_{\mathbf H_a}\!\left[
\frac{\|\widehat{\mathbf H}_a-\mathbf H_a\|_F^2}{\|\mathbf H_a\|_F^2}
\right],
\label{eq:nmse}
\end{equation}
where $\|\cdot\|_F$ denotes the Frobenius norm and the expectation is over the distribution of $\mathbf H_a$. In the federated multi-user setting, each UE keeps its encoder $\Theta_{\mathrm{en},k}$ personal and excluded from aggregation, while the decoder parameters $\Theta_{\mathrm{de}}$ are shared and
aggregated at the BS. The system model and workflow are summarized in the deployment stage section of Fig.~\ref{fig:system}.


\section{The proposed Fed-PELAD Algorithm}
\subsection{FEEL-based Personalized Encoders and a Shared Decoder}
\label{subsec:method}
We employ an autoencoder for CSI feedback, where each UE \( k \) retains its encoder parameters locally to facilitate personalized adaptation under heterogeneous data distributions. For clarity, we denote the set of parameter matrices for the decoder by \( \{W_{\mathrm{de},\ell}\}_{\ell=1}^L \), where $L$ is the number of LoRA-adapted layers and \( \ell \) indexes the \( \ell \)-th layer.
\begin{figure*}[t]
        \centering        \includegraphics[width=1\textwidth]{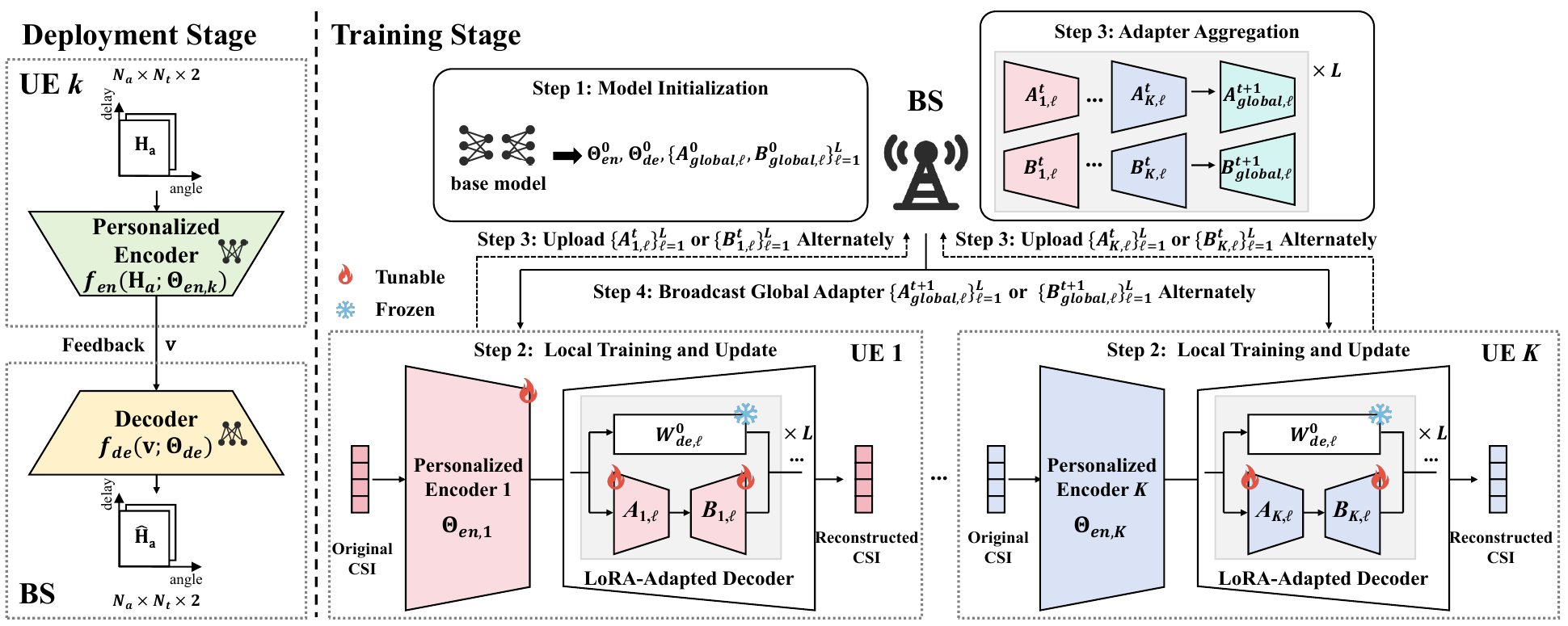}
        \caption{Fed-PELAD Framework: Personalized Encoders and a LoRA-Adapted Shared Decoder with Federated Adapter Aggregation.}
        \label{fig:system}
\end{figure*}

To reduce communication overhead, each decoder linear layer is equipped with a LoRA module. For a layer $\ell$ with frozen pretrained weights $W^{0}_{\mathrm{de},\ell}\!\in\!\mathbb{R}^{d_{1,\ell}\times d_{2,\ell}}$, we introduce low-rank adapters $A_{\ell}\!\in\!\mathbb{R}^{r\times d_{2,\ell}}$ and $B_{\ell}\!\in\!\mathbb{R}^{d_{1,\ell}\times r}$, and reparameterize the effective weights as
\begin{equation}
W_{\mathrm{de},\ell}
\;=\;
W^{0}_{\mathrm{de},\ell}
+
\Delta W_{\ell},
\qquad
\Delta  W_{\ell}
\coloneqq
\frac{\alpha}{r}B_{\ell}A_{\ell},
\label{eq:lora}
\end{equation}
where $\alpha$ is a scaling factor controlling the contribution of the LoRA update, and $r\!\ll\!\min(d_{1,\ell},d_{2,\ell})$ is the adapter rank. During training, only the LoRA adapters $\{A_{\ell}, B_{\ell}\}_{\ell=1}^L$ are trained and transmitted, while the pretrained decoder weights $\{W^{0}_{\mathrm{de},\ell}\}_{\ell=1}^L$ remain frozen throughout the process.

The overall training procedure of Fed-PELAD comprises four stages, as illustrated in the training stage of Fig.~\ref{fig:system}. 

\textbf{Step 1: Model Initialization.} A base autoencoder is first centrally pre-trained on a composite dataset. The BS then initializes personalized encoder parameters \(\Theta^{0}_{\mathrm{en}}\), the pretrained decoder weights \(\Theta^{0}_{\mathrm{de}}\), and initial global LoRA adapters \(\{A_{\mathrm{global},\ell}^{0}, B_{\mathrm{global},\ell}^{0}\}_{\ell=1}^L\) using the base model, and then broadcasts them to all UEs.

\textbf{Step 2: Local Training and Update.} Each UE $k$ performs local training on its private CSI dataset, updating its encoder parameters \(\Theta_{\mathrm{en},k}\) and local LoRA adapters \(\{A_{k,\ell}^{t}, B_{k,\ell}^{t}\}_{\ell=1}^L\) within the shared decoder. This process follows the local data flow shown in Fig.~\ref{fig:system}. After \(E\) local epochs, the updated LoRA adapters are uploaded to the BS.

\textbf{Step 3: Adapter Aggregation.} The BS aggregates the received LoRA adapters via FedAvg~\cite{mcmahan2017communication} as:
\begin{equation}
\begin{aligned}
A_{\mathrm{global},\ell}^{t+1} &= \sum_{k} \frac{n_k}{N} A_{k,\ell}^{t},
\end{aligned}
\label{eq:agg-decoder-A}
\end{equation}
\begin{equation}
\begin{aligned}
B_{\mathrm{global},\ell}^{t+1} &= \sum_{k} \frac{n_k}{N} B_{k,\ell}^{t},
\end{aligned}
\label{eq:agg-decoder-B}
\end{equation}
where $K$ is the number of participating UEs, \(n_k\) is the local data size of UE \(k\), and \(N = \sum_{k} n_k\). This corresponds to the aggregation module in Fig.~\ref{fig:system}.

\textbf{Step 4: Global Adapter Broadcast.} The BS broadcasts the aggregated global adapters \(\{A_{\mathrm{global},\ell}^{t+1}, B_{\mathrm{global},\ell}^{t+1}\}_{\ell=1}^L\) to all UEs, replacing their local versions for the next round.

Training repeats Step 2-4 until convergence or the maximum communication rounds is reached. In deployment, each UE uses its personalized encoder for compression, and the BS uses the shared decoder for reconstruction, following the deployment stage in Fig.~\ref{fig:system}.
\subsection{Alternating-Freeze Aggregation for the Decoder}
To mitigate update interference which occurs when simultaneous gradient updates of multiple LoRA adapters yield conflicting directions and unstable convergence under heterogeneous CSI data in FEEL, we introduce two complementary mechanisms for the shared decoder’s adapters: alternating freezing (AF) and learning-rate ratio calibration~\cite{koo2024towards}. AF time-multiplexes the updates of the two LoRA adapters. In round $t$, only one adapter is transmitted while the other is kept fixed: odd rounds update $B$ with $A$ frozen; even rounds update $A$ with $B$ frozen. 

The two adapters behave differently in practice: $A$ roughly sets the update direction, whereas $B$ mainly adjusts its stepsize \cite{hayou2024loraplus}. Consequently, they tend to update at different speeds. We therefore assign separate learning rates $\eta_A$, $\eta_B$ and control them via the learning-rate ratio $\eta_B/\eta_A$. The full procedure is outlined in Algorithm~\ref{alg:af-Fed-PELAD-concise}. 


\begin{algorithm}[t]
\caption{Fed-PELAD with Alternating Freeze (AF)}
\label{alg:af-Fed-PELAD-concise}
\begin{algorithmic}[1]
\Statex \hspace*{-\algorithmicindent}\textbf{Input:} Number of UEs $K$, total communication rounds $T$, and local epochs $E$.
\State BS broadcasts the pretrained weights \(\Theta^{0}_{\mathrm{en}}\), \(\Theta^{0}_{\mathrm{de}}\), and \(\{A_{\mathrm{global},\ell}^{0}, B_{\mathrm{global},\ell}^{0}\}_{\ell=1}^L\) to all the UEs
\For{$t=1,2,\ldots,T$}
  \ForAll{UE}
    \State Update $\Theta_{\mathrm{en},k}$ and $\{A_{k,\ell}^{t},B_{k,\ell}^{t}\}_{\ell=1}^L$ via minimizing \eqref{eq:nmse} for E local epochs
      \State 
      UE $k$ uploads $\{B_{k,\ell}^{t}\}_{\ell=1}^{L}$ if $t$ is odd, otherwise $\{A_{k,\ell}^{t}\}_{\ell=1}^L$.
  \EndFor
  \State \textbf{BS aggregation:}
    \If{$t$ is odd}\State Update $\{B_{global,\ell}^{t+1}\}_{\ell=1}^L$ for all $L$ layers via \eqref{eq:agg-decoder-B}
    \Else
        \State Update $\{A_{global,\ell}^{t+1}\}_{\ell=1}^L$ for all $L$ layers via \eqref{eq:agg-decoder-A}
    \EndIf
    \State BS broadcasts the update global adapter $\{B_{\text{global}, \ell}^{t+1}\}_{\ell=1}^L$ if t is odd else $\{A_{\text{global}, \ell}^{t+1}\}_{\ell=1}^L$ to all the UEs alternately
\EndFor
\Statex \hspace*{-\algorithmicindent}\textbf{Output:} The converged CSI feedback network of all UEs
\end{algorithmic}
\end{algorithm}

\section{Simulation Results}

\subsection{Simulation Settings}

\subsubsection{Datasets and Model}
The proposed Fed-PELAD framework is designed to be compatible with generic autoencoder-based CSI feedback architectures. To validate its effectiveness, we conduct experiments using TransNet~\cite{cui2022transnet}, a transformer-enhanced deep learning-based CSI feedback model with 1.11 M parameters. Following 3GPP TR~38.901, we construct a composite dataset covering \emph{UMi}, \emph{UMa}, and \emph{RMa} scenarios under both \emph{LOS/NLOS} and \emph{indoor/outdoor} conditions, resulting in 12 sub-scenarios. Channel realizations are generated with QuaDRiGa~\cite{jaeckel2014quadriga} in a sector cell of $300$\,m radius and $120^{\circ}$ angular span, with the BS located at the sector vertex. For each of the 12 sub-scenarios, $5{,}000$ static UEs are uniformly deployed, and one CSI snapshot is extracted per UE, yielding a total of $60{,}000$ samples. We first pretrain a base model via centralized learning. 
Across all scenarios, consistent radio parameters are adopted: a $4\times 8$ planar array at BS, single isotropic antenna at UE, $30$\,kHz subcarrier spacing, $52$ resource blocks, $20$\,MHz bandwidth, and $2$\,GHz carrier frequency. Scenario-specific geometries include BS heights of $10/25/35$\,m for UMi/UMa/RMa, respectively, UE height of $1.5$\,m, and minimum BS--UE distances of $10/35/35$\,m.

\subsubsection{Baselines and LoRA-adapted Layers}
To evaluate the proposed Fed-PELAD framework, we compare it under heterogeneous data scenarios against the following baselines:
FedAvg (full-model aggregation at the BS),
FedDec (FEEL-based personalized encoders and direct aggregation of all decoder parameters), Fed-PELAD w/o AF (AF optimization not being used), 
and Fed-PELAD\_Half (AF optimization with the number of communication rounds halved).
For a fair comparison of the total UE-BS transmissions, we set the number of rounds for FedAvg, FedDec, and Fed-PELAD\_Half to \(200\), and for {Fed-PELAD to \(400\). A total of $L=11$ layers of decoder are adapted with LoRA, which include all linear projections from multi-head attention modules and the remaining linear layers.

\subsubsection{Federated Training Protocol, Hyperparameters and Metrics}
We focus on the UMi scenario to evaluate the federated learning schemes. We establish five independent evaluation cohorts to enhance statistical reliability. Each cohort consists of $K=4$ UEs, obtained by randomly selecting one UE from each of the four sub-scenarios (S1–S4). The model performance is assessed under four distinct conditions, denoted as S1 to S4, which encompass all combinations of \emph{LOS/NLOS} and \emph{indoor/outdoor} propagation environments.

Each UE possesses a local dataset of 3,000 CSI samples, collected under consistent sector geometry and split in an 8:1:1 ratio for training, validation, and testing, respectively. Local training is performed using the Adam optimizer with a fixed learning rate $\eta_A=10^{-4}$, a batch size of 200, and $E=2$ local epochs per communication round. Unless otherwise specified, the following hyperparameter values are used as defaults: CR of $1/16$, LoRA rank $r=64$, coefficient $\alpha/r=5$, and a learning-rate ratio $\eta_B/\eta_A=1$. 

The final performance metric for a scenario is reported as the average result across the five corresponding cohorts. BS-side parameter aggregation follows FedAvg \cite{mcmahan2017communication} on the shared decoder LoRA parameters.
The evaluation metric for model performance is NMSE, as shown in Eq~\eqref{eq:nmse}. 

In addition, we define the per-round uplink communication load for UE $k$ as $L_{up,k}(t)$, which represents the number of parameters transmitted by UE $k$ to the BS in round $t$. 
We then introduce the Cumulative Uplink Cost (CUC) up to communication round $t$:
\begin{equation}
\mathrm{CUC}_m(t) \;=\; \sum_{\tau=1}^{t} \sum_{k=1}^{K} L_{up,k}(\tau),
\label{eq:cuc_def}
\end{equation}
which is used to assess accuracy-–overhead trade-off. For comparisons against a baseline, the relative communication uplink cost \(\mathrm{rCUC}_m\) of method \(m\) normalized by FedAvg is denoted as:
\begin{equation}
\mathrm{rCUC}_m \;=\; \frac{\mathrm{CUC}_m}{\mathrm{CUC}_{\text{FedAvg}}}.
\label{eq:rcuc}
\end{equation}


\subsection{Performance Evaluation}

\begin{table*}[t]
\centering
\caption{Per-scenario NMSE (dB) and relative communication uplink cost $\mathrm{rCUC}_m$ under UMi: Fed-PELAD vs. Baselines.}
\label{tab:Fed-PELAD-nmse}
\setlength{\tabcolsep}{7pt}        
\renewcommand{\arraystretch}{1.15} 
\begin{tabular}{lcccccc}
\toprule
\textbf{Method} & \textbf{LOS\_Outdoor (S1)} & \textbf{NLOS\_Outdoor (S2)} & \textbf{LOS\_Indoor (S3)} & \textbf{NLOS\_Indoor (S4)} & \textbf{Avg.} & \textbf{$\mathrm{rCUC}_m$} \\  
\midrule
\textbf{FedAvg}        & $-16.19 \pm 3.48$ & $-9.35 \pm 2.05$ & $-9.22 \pm 3.29$ & $-9.63 \pm 2.67$ & $-11.10 \pm 1.50$ & 100.00\% \\
\textbf{FedDec}        & $-17.28 \pm 3.77$ & $\mathbf{-10.11 \pm 2.34}$& $-9.73 \pm 3.61$ & $-10.23 \pm 2.81$& $-11.84 \pm 1.55$ & 50.85\% \\
\textbf{Fed-PELAD w/o AF} & $-18.19 \pm 4.44$ & $-9.96 \pm 2.03$ & $\mathbf{-9.75 \pm 3.28}$ & $\mathbf{-10.28 \pm 2.65}$& $-12.05 \pm 1.57$ & 42.97\% \\
\textbf{Fed-PELAD\_Half}     & $-17.05 \pm 4.06$ & $-9.07 \pm 2.14$ & $-8.84 \pm 2.92$ & $-9.37 \pm 2.28$ & $-11.08 \pm 1.48$ & $\mathbf{21.49\%}$\\
\textbf{Fed-PELAD}        & $-\mathbf{19.95 \pm 4.19}$ & $-9.82 \pm 2.16$& $-9.55 \pm 3.20$& $-10.11 \pm 2.55$& $\mathbf{-12.36 \pm 1.68}$ & 42.97\% \\
\bottomrule
\end{tabular}
\vspace{-4mm}
\end{table*}


\begin{figure}
    \centering
    \includegraphics[width=0.45\textwidth]{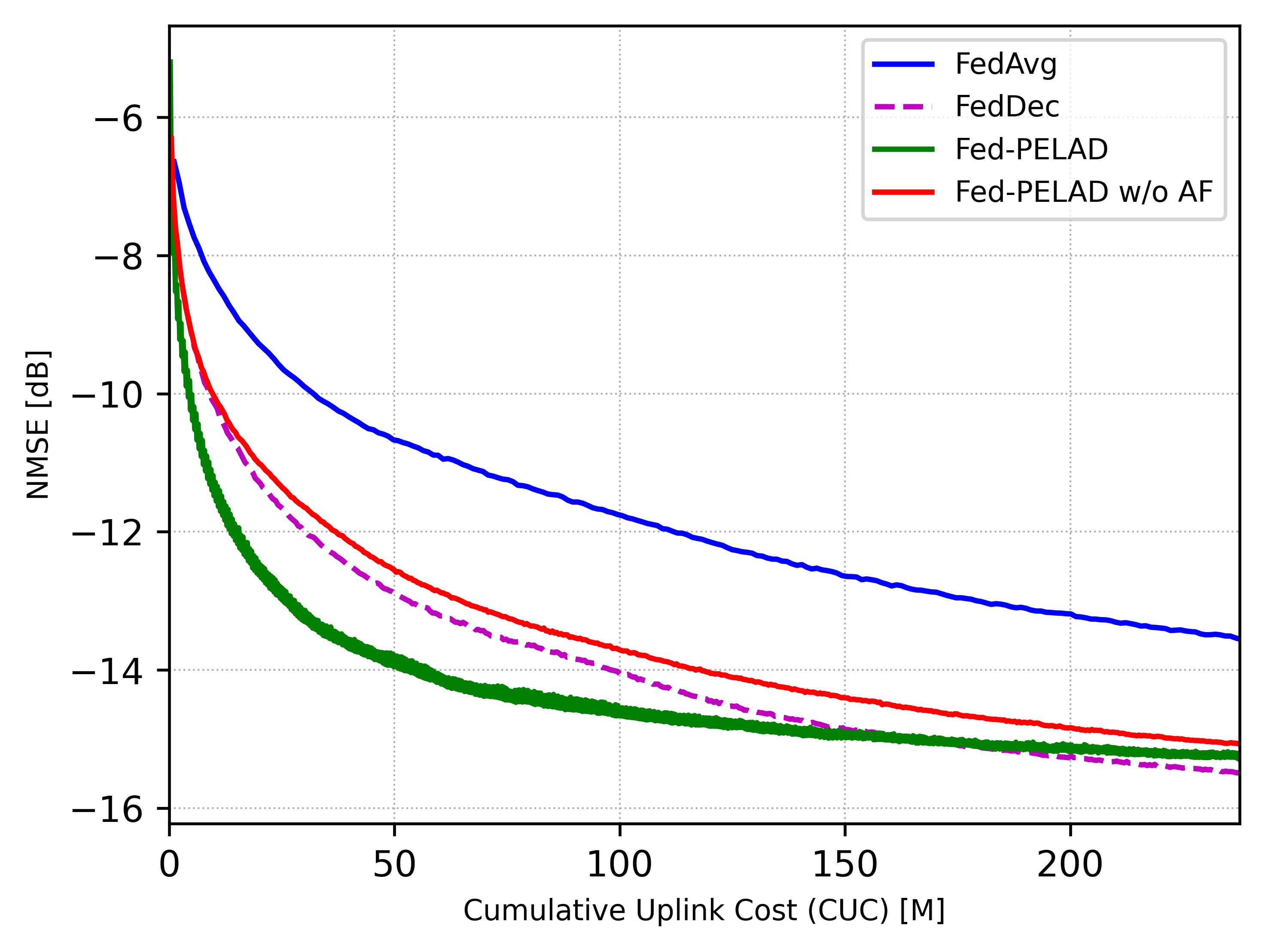}
    \caption{NMSE (dB) Performance of Fed-PELAD and Baselines vs. Cumulative Uplink Cost (Million parameters) under UMi scenario with four UEs}
    \label{fig:enter-label}
\end{figure}

Fig.~\ref{fig:enter-label} presents the average NMSE performance across four UMi sub-scenarios. As shown, Fed-PELAD consistently outperforms the baselines when CUC is below 150~M. With increasing CUC, FedDec gradually narrows the performance gap and eventually surpasses Fed-PELAD in high-resource regimes, achieving a marginal advantage of approximately 0.19~dB at the maximum communication round $T=400$. The ablation comparison with Fed-PELAD w/o AF further highlights the effectiveness of the alternating freeze mechanism, which mitigates heterogeneity-induced inconsistency in federated aggregation. As CUC increases, the performance of Fed-PELAD w/o AF approaches that of the full Fed-PELAD model. Under a halved uplink communication budget, for instance when Fed-PELAD operates at $\mathrm{CUC}=50\,\mathrm{M}$ compared with FedDec at $\mathrm{CUC}=100\,\mathrm{M}$, Fed-PELAD shows only a minor performance loss of approximately $0.27\,\mathrm{dB}$, demonstrating a superior trade-off between accuracy and uplink communication cost.


As delineated in Table~\ref{tab:Fed-PELAD-nmse}, Fed-PELAD demonstrates superior or highly competitive performance across the four heterogeneous UMi scenarios. It achieves the best NMSE in the S1 scenario and attains the highest overall average NMSE of $-12.36 \pm 1.68$ dB. Crucially, this leading performance is accomplished at a communication cost of only $42.97\%$ of FedAvg's, translating to an NMSE improvement of $1.28$ dB over FedAvg and $0.52$ dB over FedDec. 

The efficacy of the proposed alternating-freeze strategy is evident when comparing Fed-PELAD with its ablation variant (Fed-PELAD w/o AF). While both share the same low communication overhead ($42.97\%$), Fed-PELAD gains $0.31$ dB in average NMSE. Furthermore, it secures the most stable performance in the challenging S1 scenario, reducing the performance variance from $\pm 4.44$ dB to $\pm 4.19$ dB. Under an even tighter communication budget, Fed-PELAD\_Half ($21.49\%$ rCUC) remains competitive, being only $0.02$ dB worse than the full-communication FedAvg and $0.76$ dB worse than FedDec, thus achieving a better accuracy--overhead trade-off.

\subsection{Impact of Hyperparameters on Fed-PELAD}

The comprehensive hyperparameter sensitivity analysis with fixed LoRA rank $r=64$ is presented in Table~\ref{tab:hparam-comprehensive}. Specifically, a compensatory relationship is observed: as $\eta_B/\eta_A$ increases, the optimal $\alpha/r$ value decreases. For instance, at $\eta_B/\eta_A=1$, $\alpha/r=8$ achieves $-12.05$~dB, while at $\eta_B/\eta_A=10$, $\alpha/r=1$ yields $-12.08$~dB. This indicates that increasing the learning-rate ratio for adapter $B$ can effectively compensate for reducing the LoRA coefficient. The global optimum of $-12.36$~dB is attained at $(\alpha/r=1, \eta_B/\eta_A=5)$. A broader high-performance region (NMSE $< -12.00$~dB) is observed for $\alpha/r$ between 1--4 and $\eta_B/\eta_A$ between 3--7, indicating robustness within this parameter subspace.

Table~\ref{tab:hparam-r} investigates the impact of the LoRA rank $r$ under the optimal configuration ($\alpha/r\!=\!1$, $\eta_B/\eta_A\!=\!5$). The results demonstrate a clear accuracy--overhead trade-off: increasing $r$ from 2 to 64 monotonically improves the average NMSE from $-8.41$\,dB to $-12.36$\,dB, while the relative communication cost rCUC$_m$ increases from $1.34\%$ to $42.97\%$. For practical deployment, $r=32$ offers a favorable balance, achieving $-11.12$\,dB with only $21.49\%$ communication overhead, while $r=8$ provides a lightweight option ($-9.61$\,dB at $5.37\%$ rCUC$_m$) suitable for stringent bandwidth constraints.


\definecolor{best}{RGB}{200,200,200}    
\definecolor{good}{RGB}{230,230,230}    
\definecolor{mid}{RGB}{245,245,245}     
\definecolor{poor}{RGB}{255,255,255}    

\begin{table}[t]
\centering
\small
\caption{Average NMSE (dB) under Different $\alpha/r$ and $\eta_B/\eta_A$ ($r=64$)}
\label{tab:hparam-comprehensive}
\begin{adjustbox}{width=\columnwidth}
\begin{tabular}{ccccccc}
\toprule
\multirow{2}{*}{$\eta_B/\eta_A$} & \multicolumn{6}{c}{$\alpha/r$} \\
\cmidrule(lr){2-7}
 & \textbf{1} & \textbf{2} & \textbf{3} & \textbf{4} & \textbf{8} & \textbf{16} \\
\midrule
\textbf{0.5} & \cellcolor{poor}-9.92 & \cellcolor{mid}-10.99 & \cellcolor{mid}-10.66 & \cellcolor{mid}-10.83 & \cellcolor{good}-11.21 & \cellcolor{good}-11.44 \\
\textbf{1} & \cellcolor{mid}-11.00 & \cellcolor{good}-11.41 & \cellcolor{good}-11.66 & \cellcolor{good}-11.81 & \cellcolor{best}-12.05 & \cellcolor{best}-12.04 \\
\textbf{3} & \cellcolor{good}-11.66 & \cellcolor{best}-11.99 & \cellcolor{best}-12.09 & \cellcolor{best}-12.06 & \cellcolor{good}-11.94 & \cellcolor{mid}-11.53 \\
\textbf{5} & \cellcolor{best}\textbf{-12.36} & \cellcolor{best}-12.07 & \cellcolor{best}-12.03 & \cellcolor{best}-12.00 & \cellcolor{good}-11.64 & \cellcolor{mid}-11.03 \\
\textbf{7} & \cellcolor{best}-12.03 & \cellcolor{best}-12.03 & \cellcolor{good}-11.95 & \cellcolor{good}-11.86 & \cellcolor{mid}-11.34 & \cellcolor{poor}-10.58 \\
\textbf{10} & \cellcolor{best}-12.08 & \cellcolor{good}-11.99 & \cellcolor{good}-11.81 & \cellcolor{mid}-11.61 & \cellcolor{mid}-10.98 & \cellcolor{poor}-10.15 \\
\bottomrule
\end{tabular}
\end{adjustbox}
\end{table}


\begin{table}[t]
\centering
\small
\caption{Effect of LoRA rank $r$ on NMSE (dB) and relative  communication uplink cost $rCUC_m$ (fix $\alpha/r=1$, $\eta_B/\eta_A=5$).}
\label{tab:hparam-r}
\begin{adjustbox}{width=\columnwidth}
\begin{tabular}{ccccccc}
\toprule
\textbf{$r$} & \textbf{2} & \textbf{4} & \textbf{8} & \textbf{16} & \textbf{32} & \cellcolor{gray!12}\textbf{64} \\
\midrule
S1  & -14.57 & -15.12 & -15.93 & -16.75 & -17.39 & \cellcolor{gray!12}-19.95 \\
S2  & -6.47  & -6.91 & -7.49 & -8.14 & -8.98 & \cellcolor{gray!12}-9.82 \\
S3  & -6.11  & -6.52  & -7.21  & -7.89  & -8.83  & \cellcolor{gray!12}-9.55 \\
S4  & -6.51 & -7.14 & -7.81 & -8.50 & -9.28 & \cellcolor{gray!12}-10.11 \\
\textbf{Avg.} & -8.41 & -8.92 & -9.61 & -10.32 & -11.12 & \cellcolor{gray!12}-12.36 \\
\textbf{rCUC$_m$} & 1.34\% & 2.69\% & 5.37\% & 10.74\% & 21.49\% & \cellcolor{gray!12}42.97\% \\
\bottomrule
\end{tabular}
\end{adjustbox}
\end{table}

\section{Conclusion}
This paper introduced Fed-PELAD, a communication-efficient federated learning framework for CSI feedback in FDD massive MIMO systems. Fed-PELAD adopts a personalized-encoder, shared-decoder architecture, where encoder parameters remain on the UE, while a LoRA-adapted shared decoder is collaboratively trained. To enhance robustness under heterogeneous data scenarios, we further proposed an alternating-freeze schedule and calibrated the learning-rate ratio of the LoRA adapters, which jointly stabilize training and mitigate inter-UE update mismatch.
Extensive experiments under 3GPP-standard channel scenarios demonstrate the practical viability of Fed-PELAD, which substantially curtails uplink communication costs in federated training and attains an accuracy--overhead trade-off, facilitating its adoption in operational massive MIMO systems.

\printbibliography 
\vfill
\end{document}